\documentclass[10pt,leqno]{amsart}
\usepackage{algpseudocode}
\usepackage[a4paper,margin=2cm]{geometry}
\newcommand{\id}[1]{\mathit{#1}}

\begin{document}

\title[Diffusion Monte Carlo algorithm ...]{A Diffusion Monte Carlo algorithm employing depth first traversal and a stack instead of a swarm}
\author[B J Braams]{Bastiaan J. Braams}
\date{\today}
\address{Centrum Wiskunde \& Informatica (CWI), Amsterdam, The Netherlands}
\email{b.j.braams@cwi.nl}

\begin{abstract}

Diffusion Monte Carlo (DMC) and Monte Carlo for particle transport with importance sampling both involve simulations of weighted walkers that undergo birth and death processes (splitting and Russian Roulette).
The established implementations of these methods are quite different: Particle simulation Monte Carlo employs a stack to handle the splitting history whereas in traditional DMC one follows a swarm of walkers.
The particle simulation Monte Carlo approach involves a depth first traversal of the visited configurations whereas the traditional DMC approach may be seen as a breadth first traversal.
In the present work the implementation of a depth first, stack based approach to DMC is described and a complete code is presented.
The depth first approach, called DMCD here, can be more memory efficient than the breadth first approach, both for total memory and for use of a memory hierarchy and of co-processors.
The implementation appears very natural for population control and for descendant weighting and it unifies algorithmic treatment of the eigenvalue problem (DMC) with the linear equation problem (particle transport).
A concern with DMCD that is not present in the breadth first approach, and that is successfully addressed here, is the need to maintain a pool of starters for use when a new walker is required and the stack is empty.
The DMCD approach appears to have the potential to become the preferred implementation for many DMC applications.

\end{abstract}

\maketitle

\section*{Introduction}

The history of a weighted random walk Monte Carlo simulation with splitting and Russian Roulette (branching, birth and death, cloning and killing) can be represented by a forest (a collection of trees).
Each node is associated with a walker (also called a replica or a particle) at a particular time or iteration number.
The node carries data that we call the configuration; what this means depends on the application.
Each node also carries a weight, and in many applications this would be a nonnegative real number although more general weights are possible.
Finally the time or iteration number may be carried along as a distinct node property.
The roots of the trees represent walkers freshly created from some initial distribution, leaves represent the final state of a walker that is terminated (killed), nodes with one edge in and one edge out are ordinary steps in the evolution, and nodes with multiple edges out represent proper splitting events.

The techniques of splitting and Russian Roulette were introduced in connection with Monte Carlo methods for particle transport already in the 1940s \cite{Kahn1954,CC1974,Brie2000,Spa2009,RF2013,LK2018,CHKW2022}.
These techniques were adopted for Diffusion Monte Carlo (DMC) and the related Green Function Monte Carlo for sampling from a solution of the Schrödinger equation starting in the 1960s.
The book \cite{And2007} offers a collection of summaries of the early and more recent literature.
I note the foundational papers and reviews \cite{SW1991,UNR1993,NU1998,FMNR2001,NTD2010,AZL2012,AAZ2024,DRFB2025}.
These techniques are also central in Sequential Monte Carlo (particle filtering) for sampling from a sequence of distributions,
especially in the context of data assimilation \cite{DGA2000,DMDJ2006,CGM2007,ADH2010,NLS2019,CP2020,DHJW2022,FNOS2025,SSDKJ2025}.

From a mathematical or numerical analysis perspective all these methods belong to stochastic linear algebra for situations where the solution vector is only sampled.
This perspective is prominent in \cite{HW2014,LW2017,Web2019}.
The particle transport simulations then appear as a linear equation solver, DMC is used to obtain a dominant eigenvector, and Sequential Monte Carlo is a certain nonstationary iteration.
In the case of DMC an eigenvalue has to be determined along with the eigenvector and in the case of Sequential Monte Carlo a sequence of normalization constants (partition functions) is obtained.

In particle or radiation transport simulations via Markov Chain Monte Carlo with importance sampling it is completely standard to walk through the tree or forest of configurations depth first and to manage the splitting events using a stack.
(This is for a sequential implementation. Parallelism can be introduced at many levels and the proper choice is too much problem dependent to be considered here.)

In DMC the need to determine also an eigenvalue has led to a different kind of implementation.
The basis is a swarm of weighted walkers all initialized at the same simulation time and advancing (conceptually) in parallel, so via a breadth first traversal of the tree or the forest.
At selected times in the simulation the weights are rebalanced by splitting and Russian Roulette.
The rates of these processes are controlled by the unknown eigenvalue, and the estimated value is adjusted in  order to keep the swarm size approximately constant.

The present manuscript describes the development and a demonstration implementation of a depth first, stack based approach to DMC.
I call the approach DMCD here, and contrast it with the established breadth first, swarm based approach that will be called DMCB on occasion.
The implementation is in a Fortran (2018/2023) code that is included as a complete package in ancillary information with this manuscript.

For the concerns of the present work Sequential Monte Carlo is very similar to DMC, and it will be left out of the further discussion.
Also left out of the further discussion will be all the very important application oriented concerns of DMC, such as accurate time discretization, accurate nodal surfaces and accurate guiding functions.
We are concerned here with the implementation of DMC as a stochastic eigensolver, taking the underlying operator and the associated weighted Markov process for granted.

Important issues for DMC that are independent of the underlying operator include population control and associated bias \cite{ACK2000,BR2004,WH2006} and descendant weighting (also called pure sampling or forward walking) \cite{SW1991,BRL1991,BS1998,OR2015}.
Those issues are addressed here from the perspective of the implementation of DMCD.
The issues are addressed from the established DMCB perspective in the descriptions of widely used DMC codes \cite{DPS2025,NTD2020,KAB2020,NAB2020,WPK2023,SGD2026}.

The idea to use a depth first stack based implementation of DMC may not be entirely original.
A suggestion to use splitting in single walker trajectories in DMC appears in a report \cite{TREX2022} from the TREX project (Section 3, adding a branching step to PDMC), but the suggestion did not make it into the project publications such as \cite{SGD2026}.
Perhaps the idea has also been considered from the particle simulation side for expanding into DMC, but I have not seen it.

The next Section introduces the algorithmic considerations that go into the present DMCD demonstration code, especially with respect to managing the particle balance, sources of bias (finite stack size versus finite swarm size), descendant weighting and maintaining a pool of starters for newly initiated walks.

The proper description of the code follows in one Section on all the peripheral modules and one Section on the DMCD main module.
Module data, types defined in the modules and their principal type bound procedures are described there, sometimes using pseudo-code.
This is intended as an adequate introduction to the actual code in ancillary information, which is well commented.

This manuscript is not about applications, but the Section Experiments describes a few simulations on a model system to illustrate the promise of the approach.
This is followed by the Conclusions and by a guide to the Ancillary information.

In closing the Introduction I note what is potentially of interest here.

For practical DMC I expect that the main advantage of DMCD will lie in more efficient memory use, including more efficient use of a memory hierarchy and of co-processors.
Although many things need to be validated on realistic applications, it appears that the stack in DMCD can be much smaller than the swarm in DMCB.
This can be valuable for applications of DMC in which a walker configuration is a very large object or in which, in a parallel program, many DMC calculations must run independently.

For applications in which walkers are not very large it can be advantageous that in DMCD a processor will work on a single walker for many steps (until splitting or termination), allowing a more efficient use of a cache or other memory hierarchy and of co-processors.

Another consideration is that descendant weighting has a very natural implementation (the required history information is right there in the stack) and it may be more accurate in DMCD than in DMCB due to some natural averaging over walker trajectories between splitting events (as discussed later in the manuscript).
Furthermore, the unification of algorithms for particle transport Monte Carlo and for DMC can be of independent merit.

The one complication that is present in DMCD that is not present in the breadth first approach is the need to maintain a pool of walkers that may be used as starters when a trajectory is terminated and the stack is empty.
This task is addressed in what appears to be an entirely satisfactory manner.

\section*{Algorithmic considerations}

As one thinks more closely about a depth first, stack based approach (DMCD) to DMC it becomes clear that some issues are very different in DMCD than in the breadth first, swarm based approach (DMCB).
In this section these issues are discussed in an informal way.
The code description in the subsequent two sections is more precise, and the code in the ancillary information is authoritative.

\subsection*{Population control and particle balance}

Population control is necessary due to the competing processes of splitting and Russian Roulette, which must be kept in balance.
The population control is very different and, I find, much simpler in the DMCD approach than in DMCB.
A stack is naturally of variable size and therefore the DMCB concern to keep the swarm size constant or nearly constant is absent in DMCD.
The stack size may fluctuate wildly over a calculation.
This poses no difficulty; population control on every timestep works well.
By contrast, in DMCB weight rebalancing is a major event that one does not want to do on every timestep.

In the implementation here this population control for DMCD is done by rescaling the weight of the active walker on each timestep.
For this purpose we accumulate (over all iterations) the walker weights just before and just after a timestep in quantities $\id{wt0a}$ and $\id{wta}$.
(This involves a smoothed exponentially decaying weight as will be described later.)
Then the rescaling factor is just $\id{wt0a}/\id{wta}$.

(In earlier explorations of DMCD I dynamically adjusted the rates of splitting and termination in order to keep the stack about half full at all times and in any case to prevent it from either filling up completely or from going empty.
The present simple weight control is much cleaner, and we deal with the stack filling up or going empty as will be described.)

\subsection*{Bias due to finite stack size}

The finite size of the swarm in the DMCB approach is a source of bias in the calculation.
The finite size of the stack in the DMCD approach also creates bias, but in a different way.

If the criteria are met to split the active walker, but the stack is full, then we have a problem.
One may think to just continue the walk then without splitting.
That policy is free of bias, but it can create very large variance.
My choice is to continue the walk without splitting, but to limit the walker weight to some (tunable) value at least as large as the threshold for splitting.
That keeps the variance in bounds, but it creates a bias whose magnitude needs to be characterized.
(I expect it to decay inversely with the maximum stack size in typical situations, that being the natural rate of decay for how often the stack is full.)
For a model problem the bias is very small, as discussed in the Experiments section.

A potential source of bias associated with the stack going empty is addressed below under Starter bias.

\subsection*{Descendant weighting}

Descendant weighting requires history information, and in the stack based approach history information is right there in the stack.
The question is how to use that information.

For descendant weighting some notion of distance in time or in iteration number is needed; quantities of interest (QoI) at one time are accumulated weighted by walker weights at a later time.
The suitable distance in iteration number is clearly problem dependent: if a calculation is changed by halving the timestep then one may expect that the suitable distance in iteration number for descendant weighting is doubled.
On the other hand, the distance in number of intervening splitting events will be largely insensitive to changes in the timestep.
In the stack based approach it is therefore both natural and attractive to measure distance by number of splitting events, i.e., by the distance in the stack.

Notice now that, when measuring distance by number of intervening splitting events, an entire sequence of walker steps has identical distance to the present configuration.
Therefore, in the DMCD implementation of descendant weighting we accumulate both the QoI and the weights over a complete sequence of steps from walker creation (in a splitting or initiation event) to walker split or termination.
These accumulated weights and QoI become part of the walker information.
(See the description of the $\id{WkHist\_Type}$ data type in the code description below.)
If a variable timestep is used then this accumulation within a sequence of steps must be weighted by the timestep.

\subsection*{Starters and starter bias I: The stack and starter pool as one cyclic array}

The role and influence of starter configurations for a walker trajectory is completely different and more complicated in DMCD than in DMCB.
For either implementation the weighted Markov process must be suitably mixing and we take that for granted here.

In the swarm based DMCB one initializes the swarm in some way.
In the subsequent evolution walkers may die and walkers may split, but there is no need for new starters after initialization of the swarm.
In the most common situation, ultimately the entire swarm is descended from one starter and the starter bias decays inversely to the simulation time.

In the stack based DMCD as implemented here it is natural for the stack to go empty.
If the active walker is terminated (through Russian Roulette) at such a time then a new starter is required.

So one needs a pool of starters, but we must not assume that such a pool is provided from outside the algorithm.
Instead, the pool of starters has to be created and maintained in the natural course of the iterations.
This is not a simple matter, but the treatment in the present code works very well for the model problem.
The treatment will be motivated and described quite informally in the following few paragraphs; then a bit more formally in the section on Module $\id{DMCD}$, procedure $\id{Strt}$, and the precise description is the code in the ancillary material.

Preliminary to the design of the starter pool and its maintenance procedures, note that a new starter is only needed if the stack is empty.
Heuristically, if the stack is getting low then we like to have many starters in the pool.
If the stack is close to full then we can do with fewer starters; the starter pool will be improved as the iterations progress before the stack is empty.

The stack is implemented as a fixed size array.
We may be tight for space, and it looks feasible and not unnatural to use the same fixed size array for the starter pool, which then occupies the locations complementary to those for the stack.
If a walker is drawn from the starter pool then it must be removed from that pool; starters that are used should be independent as much as is possible.

This leads to the design chosen in the DMCD module to implement the stack and starters pool as one cyclic array.
All indices pointing into that array are to be understood modulo the array size.
The index pointing to the bottom of the stack becomes a dynamic variable, together with the size of the stack.

Let us formalize this.

We have array $\id{wks}(0:\id{nwks}-1)$ (Fortran conventions, not Python, so the size of the array is $\id{nwks}$, which is a fixed integer).
From here on we understand the array $\id{wks}(:)$ as a cyclic array, indices into the array understood modulo $\id{nwks}$, and the array being used for both the stack and, in the complementary positions, the starter pool.
We have dynamic integer variables $\id{iwk0}$ and $\id{nwk0}$ that are respectively the index in $\id{wks}(:)$ for the base of the stack and the number of elements in the stack.
Therefore the stack elements are $\id{wks}('is')$ for $\id{iwk0}\leq is<\id{iwk0}+\id{nwk0}$ and the starter pool elements are $\id{wks}('is')$ for $\id{iwk0}+\id{nwk0}\leq is<\id{iwk0}+\id{nwks}$.
Here and elsewhere in the informal description and in code comments, $\id{wks}('is')$ stands for $\id{wks}(\hbox{modulo}(is,\id{nwks}))$.

\subsection*{Starters and starter bias II: Maintaining the starter pool}

We discuss now how the (dynamic) starter pool section, $\id{wks}('is')$ for $\id{iwk0}+\id{nwk0}\leq is<\id{iwk0}+\id{nwks}$, is to be maintained.

Informally the treatment starts with the idea that a walker in the starter pool, say $wk'$, has a quality index $\id{l0}$ that represents its degree of randomness as a representative element of the history of the iterations.
If a new slot opens up in the starter pool (because a new active walker is taken from the stack), then the current active walker can be copied into $wk'$, but this has a very low quality of randomness.
We assign it an index $\id{l0}=0$.
In general, at any time during the iterations, if $wk'$ has quality index $\id{l0}$ then we may replace it with a copy of the current active walker with probability $2^{-\id{l0}-1}$, and if we make that replacement then assign it the new quality index $\id{l0}+1$.
That is the idea for a pool of one, but we need a larger pool.

The top element of the stack, if the stack is not empty, has index $is=\id{iwk0}+\id{nwk0}-1$.
If a new walker is needed from the stack then that $\id{wks}('is')$ is used and the variable $is$ is decremented by 1.
In the algorithm introduced here the same operation is used if the stack is empty.
Therefore the highest quality starters (in the sense introduced above) should be found at the top of the starter pool section of $\id{wks}(:)$, that is just below the bottom of the stack section of $\id{wks}(:)$.
That is fine, because these elements are the most stable members of the starter pool.
Elements on the other side, near the bottom of the starter pool or just above the top of the stack, are volatile; they can disappear from the starter pool shortly due to fluctuations in the stack size.

To formalize the procedure further we need to specify a data structure for those quality indices $\id{l0}$ for the starter pool section of the $\id{wks}(:)$ array.
For that purpose I introduce an integer array $\id{isql}(0:\id{nsql}-1)$ of size $\id{nsql}$ (to be specified).
This array will be ascending ($is\leq js\Rightarrow\id{isql}(is)\leq\id{isql}(js)$) and we also maintain $\id{iwk0}+\id{nwk0}\leq\id{isql}(:)\leq\id{iwk0}+\id{nwks}$.
It means that all elements of $\id{isql}$ point into the starter pool portion of $\id{wks}(:)$ or to the element just above the pool.
The invariants needs attention whenever $\id{iwk0}$ is decreased (that is when a walker is drawn from the top of the starter pool) and whenever $\id{nwk0}$ is increased (which happens at a splitting event).

The interpretation of $\id{isql}(:)$ is that elements $\id{wks}('is')$ for which $\id{isql}(\id{l0})\leq is<\id{isql}(\id{l0}+1)$ have quality $\id{l0}$.
(Here, $\id{isql}(\id{nsql})$ is understood as $\id{iwk0}+\id{nwks}$.)
Any elements $is:is<\id{isql}(0)$ have quality $(-1)$.

Now comes a messy point.
At the time of the original (v1) Arxiv submission of this manuscript the procedure operated as follows.
At each timestep a value $\id{l0}$ is drawn from a density decreasing as $2^{-\id{l0}-1}$ and it is checked if there is a starter pool element that has quality less than this $\id{l0}$.
If so then among such elements the one with the highest index $is$ is replaced by a copy of the present active walker and $\id{isql}(:)$ is adjusted.

It was found that this could randomly lead to a collapse of the calculation in the following way.
The very first walker makes a step to a state of lower weight, below the threshold for Russian Roulette, and it gets killed.
It is, however, copied to the pool of starters, where it becomes the lead starter, and so the new walker is just the same as the one just killed, with the reduced weight.
This walker also makes a step to a region of low weight and gets killed.
After a few such steps there is no way out.
The process is all the time continuing with the same walker, with a weight that is decreasing even down to the underflow limit.

The fix for that issue is not complicated.
If the active walker has a weight below the base weight then the copy that would go into the starter pool is subjected to Russian Roulette.
It may go to the starter pool (if the other conditions are satisfied) only if it survives the Russian Roulette, and then its weight is increased to the base weight.
Pseudo-code for this is provided in the description of procedure $\id{Strt}$ in module $\id{DMCD}$.

\section*{Code description I: Peripheral modules}

The present section describes modules that are either completely independent of the DMC application or that are loosely connected to it but still independently viable.
The reader may want to glance over this section to know where to look when reference is made to some peripheral code within the description of the DMCD module proper.

\subsection*{Modules $\id{Core}$, $\id{Random}$ and $\id{IPS}$}

These three modules contain data and procedures that are entirely independent of the DMC application.

The (Fortran) kind parameters for the code are in $\id{Core}$.
A procedure $Core\_Assert$ triggers an error stop if its argument is $False$ while $Core\_Expect$ triggers a warning message under that condition.
A few other routines from the $\id{Core}$ module find isolated use in this program.

Procedure $\id{Random\_True}$ with real argument $r$ returns $\id{True}$ with probability $median(0,r,1)$.
Procedure $\id{Random\_Gaussian}$ returns a real value or an array of reals drawn from the standard normal distribution. These procedures rely on the Fortran $\id{random\_number}$ intrinsic.

The $\id{IPS}$ module maintains a collection of name-value pairs.
Its intended use is for internal code parameters.
The name-value pairs are read from an input file at the start of execution.
They are then accessible by calls such as $\id{ival}=\id{IPS\_GetI}('name',\id{ival0})$ to return the integer value associated with the name $'name'$, or to return $\id{ival0}$ if the input file did not supply a name-value pair for $'name'$.

\subsection*{Module $\id{Sys}$}

Module $\id{Sys}$ defines an extensible type $\id{Sys\_Config\_Type}$ that provides the public interface to procedures specifying the physical or mathematical system for the DMC simulation.

The base type $\id{Sys\_Config\_Type}$ contains only an integer categorical variable $ic$ that is envisaged to have a small range $0\leq ic<ncat$.
In interesting extensions of the base module the extended type will contain geometric or other data describing a configuration.
(The base type with only the one integer variable could be used to implement some stochastic linear algebra for testing and development purposes, but I have not pursued that.)

The main public procedures acting on an object of class $\id{Sys\_Config\_Type}$ are $\id{Init}$, $\id{Step}$, $\id{Quan}$, $\id{Wgtb}$ and $\id{ICat}$.
Procedures bound to $\id{Sys\_Config\_Type}$ are intended to serve as the public interface also for more interesting extensions of the type and module.

Procedure $\id{Init}$: Provide a configuration that may be used as a starter for the DMC calculation; i.e.\ a starting configuration for a walker.
In normal intended use this can be any typical configuration.
In a variation that is introduced for testing and development purposes the configuration is drawn from a special distribution.
Ideally (for testing purposes) this would be the stationary distribution of the DMC process.
(This is indeed achieved in the $\id{SysGaus}$ module that refines $\id{Sys}$ and that is described below.)

Procedure $\id{Step}$: Perform one iteration or timestep on the system.
Input is a state $\id{cf}$ and the procedure returns probabilistically a new state $\id{cf}'$ and a multiplicative weight change $rw$.
The actual weight of any configuration (or walker) is to be maintained in the calling procedure.

Procedure $\id{Quan}$: Evaluate QoI of a configuration.
Input is a configuration $\id{cf}$ and the procedure returns a real vector $qv$.
The calling procedure may maintain accumulated weighted QoI, possibly categorized by the value of the $ic$ component of the configuration $\id{cf}$.

Procedure $\id{Wgtb}$: Input is a configuration $\id{cf}$ and the procedure returns a base weight $\id{wtb}$ for the configuration for use in importance sampling.
The expected use of $\id{wtb}$ in the calling procedure is that a walker with weight $\id{wt}$ is liable to be terminated by Russian Roulette if $\id{wt}/\id{wtb}<1$ and may be split if $\id{wt}/\id{wtb}$ is at or above some threshold.

Procedure $\id{ICat}$: Returns the value of the $ic$ categorical variable.

\subsection*{Module $\id{SysGaus}$}

In module $\id{SysGaus}$ the type $\id{Sys\_Config\_Type}$ is extended to $\id{SysGaus\_Config\_Type}$ to provide operations for an analytically tractable parameterized model system for which the state is described by a real vector and for which the stationary distribution of states is a centered Gaussian with known variance.

The model involves an integer parameter $nd$ and real parameters $a0$, $a1$, $a2$, $a3$ and $a4$.
These and some derived quantities are constant over a calculation.
They are module data and are not a part of the type.

Type $\id{SysGaus\_Config\_Type}$ contains in addition to the categorical variable $ic$ from the parent type a real vector $x$ of dimension $nd$ that describes the state of the system.
As implemented here the variable $ic$ takes only the value $0$; the action is in $x$.

Procedure $\id{Init}$: $ic$ is set to 0 and normally $x$ is set to 0.
In a variation for testing and development, $x$ is drawn from the known stationary distribution.

Procedure $\id{Step}$: The $\id{Step}$ procedure for $\id{SysGaus\_Config\_Type}$ implements a linear Gaussian model chosen for analytical tractability.
For input configuration characterized by $x$ the procedure returns $x'$ and the multiplicative weight change $rw$.
\begin{align*}
x'&=\sqrt{a0}\cdot x+\sqrt{a1}\cdot u\\
rw&=\exp(x^Tx/(2a2)-x'^Tx'/(2a3))
\end{align*}
in which $u$ is drawn from the standard (zero mean, unit variance) normal distribution.

Constraints include $0\leq a0$, $0<a1$, $a2\not=0$ and $a3\not=0$.
There are further constraints to ensure that the evolution has a well defined stationary distribution, which is then a zero-mean Gaussian.
The variance of the stationary distribution (in each dimension) both for regular weighting and for descendant weighting is calculated in the code in dependence on the parameters $a0$ through $a3$.

Procedure $\id{Quan}$: The QoI are the lowest order moments of $x$.

Procedure $\id{Wgtb}$: The $\id{Wgtb}$ procedure involves the parameter $a4$ and it returns $\exp(x^Tx/(2a4))$.
Parameter $a4$ must be nonzero, but it can have either sign.

\subsection*{Modules $\id{Walker}$ and $\id{WkHist}$}

Module $\id{Walker}$ defines a type $\id{Walker\_Type}$ and implements basic operations on that type for weighted walkers (particles, replicas) in a Monte Carlo simulation.
The module depends on the module $\id{Sys}$ or an extension of $\id{Sys}$ to define the physical system via the type or class $\id{Sys\_Config\_Type}$ and operations on that class.

An object of type $\id{Walker\_Type}$ contains a $\id{Sys\_Config\_Type}$ object $\id{cf}$ together with a walker weight $\id{wt}$ and a base weight $\id{wtb}$.
These quantities were already introduced in the description of the modules $\id{Sys}$ and $\id{SysGaus}$.

The procedures bound to $\id{Walker\_Type}$ are all entirely straightforward, building on procedures bound to $\id{Sys\_Config\_Type}$.
I note here only one such procedure.

Procedure $\id{Step}$: Input is a walker $\id{wk}$ with components $\id{cf}$, $\id{wt}$ and $\id{wtb}$.
A call to the $\id{Step}$ routine of $\id{Sys\_Config\_Type}$ returns $\id{cf}'$ and $rw$.
Then $\id{wt}'\gets rw\cdot\id{wt}$ and $\id{wtb}'$ is obtained from the $\id{Wgtb}$ routine for $\id{Sys\_Config\_Type}$.

Module $\id{WkHist}$ provides a type $\id{WkHist\_Type}$ that extends $\id{Walker\_Type}$ to describe walkers that carry some history. The intended use is for descendant weighting in our mode of DMC calculation.
Specifically the $\id{WkHist\_Type}$ contains additional fields $wti$, $dti$ and $\id{qvi}$.

Real variable $\id{wti}$ is the summed weight over time for this walker, weighted by timestep, since its creation in a splitting event or as a newly initialized walker.
(In the present code the timestep is constant and equal to 1, so $\id{wti}$ is really just a summed weight.)

Real variable $\id{dti}$ is the summed timestep for this walker since its creation as for $\id{wti}$.

Real array $\id{qvi}(0:nqoi-1,0:ncat-1)$ are the summed QoI for this walker, by category and weighted by timestep, since its creation as for $\id{wti}$.
The first dimension in $\id{qvi}$ corresponds to the dimension of the QoI vector. The second dimension is for the categorical variable $ic$; so $\id{qvi}(:,ic)$ is the summed and weighted QoI vector with summation restricted to configurations in category $ic$.
The summation is over the life of that walker since its creation event.
($\id{qvi}(:,ic)$ will be $0$ for categories never visited.)

The procedures bound to $\id{WkHist\_Type}$ are all straightforward.
The new quantities $\id{wti}$, $\id{dti}$ and $\id{qvi}$ are accumulated in the $\id{Step}$ procedure that overrides the $\id{Step}$ procedure bound to $\id{Walker\_Type}$.

\subsection*{Modules $Main$ and $MainTriv$}

These modules support the main program for the DMC calculation, but in a completely trivial way.

Module $Main$ provides an abstract framework in which a calculation is a sequence of $\id{Init}$ (initialize everything), $Calc$ (carry out a main loop) and $Done$ (wrap up the calculation).

Module $MainTriv$ is the trivial concrete version of $Main$ in which the iteration in $Calc$ is controlled by only an iteration counter.
The iteration is carried out for a fixed number of steps with diagnostic printing at fixed intervals.

\section*{Code description II: DMCD main module}

Module $\id{DMCD}$ implements the new depth first, stack based approach to DMC calculations.
The module defines a type $\id{DMCD\_Type}$ and operations on that type.
It also contains some module data (variables declared at the module level, outside any procedure) that are constants for the calculation.
Module $\id{DMCD}$ uses the support modules $\id{Core}$, $\id{Random}$ and $\id{IPS}$, the application specific module $\id{Sys}$ (or an extension of $\id{Sys}$) and the module $\id{WkHist}$ for walkers with a history.
In fact, procedures from $\id{Sys}$ are only invoked in an initialization routine.
After initialization, all the system-specific calculations are done through procedures from the $\id{WkHist}$ module.
(A walker $\id{wk}$ of type $\id{WkHist\_Type}$ contains a configuration $\id{cf}$ of class $\id{Sys\_Config\_Type}$ and the system-specific calculations are in procedures bound to the type of $\id{cf}$.)

\subsection*{Data fields in $\id{DMCD\_Type}$}

The data fields in $\id{DMCD\_Type}$ are dynamic quantities describing the state of a DMCD calculation.

$\id{wk}$, of type $\id{WkHist\_Type}$. At any time $\id{wk}$ is the active walker.

$\id{wks}(0:\id{nwks}-1)$, an array of type $\id{WkHist\_Type}$. This array contains the stack of walkers that forms the parentage (associated with the splitting operation) of the active walker.
The part of the array that is complementary to the (variable size) stack contains the dynamic pool of walkers to be used as starters whenever a new active walker is needed and the stack is empty.

$\id{wt0a}$, $\id{wta}$, real of (Fortran) $\id{kind}=\id{WP2}$ (additional precision).
$\id{wt0a}$ is the accumulated walker weight (using time-varying exponential weighting as described further down) of the walker weight just before a step.
$\id{wta}$ is the accumulated walker weight in the same way just after a step.
These weights are accumulated over the entire history of the simulation.
The ratio $\id{wta}/\id{wt0a}$ is used for weight rescaling after each step to maintain approximately constant expected walker weight.

$\id{qa}(0:nqoi-1,0:ncat-1)$, real ($\id{kind}=\id{WP2}$) array.
For $0\leq ic<ncat$, $\id{qa}(:,ic)$ is the accumulated regular weighted QoI (with exponential weighting as for $\id{wt0a}$ and $\id{wta}$) for walkers in category $ic$.
(To be clear, the category variable $ic$ of a walker is a dynamic quantity.
At each timestep the QoI of a walker is accumulated into the column of $\id{qa}$ that corresponds to its category at that timestep.)

$\id{qda}(0:nqoi-1,0:ncat-1)$, real ($\id{kind}=\id{WP2}$) array.
For $0\leq ic<ncat$, $\id{qda}(:,ic)$ is the accumulated descendant weighted QoI for walkers in category $ic$.
Different from $\id{qa}(:,:)$ there is no decay in the weights used for accumulating $\id{qda}$.

$\id{iwk0}$, $\id{nwk0}$, integer.
These variables delimit the stack and the starter pool portion of $\id{wks}(0:\id{nwks}-1)$.
Invariant: $\id{iwk0}+\id{nwk0}\leq\id{nwks}$ and $0\leq\id{nwk0}\leq\id{nwks}$, where $\id{nwks}=size(\id{wks}(:))$.
$\id{wks}(:)$ is viewed as a cyclic array and in the comments we denote by $\id{wks}('is')$ the element $\id{wks}(modulo(is,\id{nwks}))$.
For $\id{iwk0}\leq is<\id{iwk0}+\id{nwk0}$, $\id{wks}('is')$ is a stack walker.
For $\id{iwk0}+\id{nwk0}\leq is<\id{iwk0}+\id{nwks}$, $\id{wks}('is')$ is a starter pool walker.
Either the stack or the starter pool may be empty and
their combined size is $\id{nwks}$.

$\id{isql}(0:\id{nsql}-1)$ where $\id{nsql}$ is the number of binary digits in the default real type.
(I expect $\id{nsql}=53$ for large applications, but $\id{nsql}=24$ is possible too.)
The array $\id{isql}(:)$ is used in the $\id{Strt}$ procedure for maintenance of the starter pool portion of $\id{wks}(:)$.
See the $\id{Strt}$ procedure for the algorithm.

$\id{ispf}$, integer.
This variable counts the number of steps of the active walker since the most recent restart event.
Drawing a walker from the stack or (if the stack is empty) from the starter pool is a restart event.
Splitting the active walker is a restart event for the child that continues as the active walker.

\subsection*{Module data in $\id{DMCD}$}

According to general policy for this code, module data are constants for the calculation.
They may be initialized through the $\id{IPS}$ module and then they remain unchanged.
In the description here I do not distinguish between module data and submodule data.

$\id{nwks}$, integer.
The size of the $\id{wks}(:)$ array used for the stack and the starter pool.
The default value in the code is $\id{nwks}=256$.
In the experiments for the $\id{SysGaus}$ model system this is large enough so that there is no measurable bias at the resolution of these runs.

$\id{nwkd}$, integer.
The distance in the stack (distance in walker parentage) that is used for descendant weighting.
This parameter is used in the $\id{Desc}$ procedure that is called from $\id{PBal}$.
The default value in the code is $\id{nwkd}=8$.

$\id{nspf}$, integer.
The maximum number of steps that is allowed before a splitting or termination event is forced.
(Such a bound is needed to ensure that a stack history is obtained even if the $\id{Step}$ procedure exactly preserves a walker's weight ratio $\id{wt}/\id{wtb}$.)
The default value in the code is $2^{16}$, which is effectively infinite in a calculation in which splitting or termination are not rare events, but which is still small compared to the total number of steps in the calculation.

$\id{isrc}$, integer.
An indicator variable to say if a starter pool is used ($\id{isrc}=0$, the intended normal case) or if starters are drawn from some independently specified distribution ($\id{isrc}\not=0$, for some testing and development purposes).

$\id{rsp0}$, real, $2\leq\id{rsp0}$, default value $8$.
Unless the stack is full, split the active walker if $\id{rsp0}\leq\id{wt}/\id{wtb}$.

$\id{rsp1}$, real, $\id{rsp0}\leq\id{rsp1}$, default value $16$.
If the stack is full, truncate the weight of the active walker at each step so that $\id{wt}/\id{wtb}\leq\id{rsp1}$.
(This is a variance control measure as described under the $\id{PBal}$ procedure.)

$\id{rdec}$, real, nonnegative; default value $0.5$.
$\id{rdec}$ controls the decay rate for the time-varying exponential weighting in the accumulation of weights and QoI.
This is described with the procedures $\id{WDec}$ and $\id{Step}$.

\subsection*{Principal procedures bound to $\id{DMCD\_Type}$}

I describe here the driver procedure $\id{Adva}$, its principal work procedures $\id{WDec}$, $\id{Step}$, $\id{Strt}$ and $\id{PBal}$, and a few procedures called by $\id{PBal}$.

\subsection*{Procedure $\id{Adva}$}

Procedure $\id{Adva}$ advances the calculation one iteration or timestep.
This is the main procedure (besides an initialization and a diagnostic printing procedure not described here) by which module $\id{DMCD}$ is used from outside the module.
The action in $\id{Adva}$ has four components that are executed sequentially on every call.
\begin{algorithmic}
\State $\id{WDec}$ [Calculate the decay parameter for history weighting.]
\State $\id{Step}$ [Advance the active walker one iteration or timestep; maintain accumulated weights and QoI.]
\State $\id{Strt}$ [Probabilistically enter the active walker into the pool of starters.]
\State $\id{PBal}$ [Manage the particle balance by Russian Roulette and Splitting.
Descendant weighting is done here too.]
\end{algorithmic}

\subsection*{Procedure $\id{WDec}$}

Procedure $\id{WDec}$ returns $r0=\exp(-\id{rdec}/(icnt+1))-1$.
Parameter $\id{rdec}$ is $O(1)$ (default value $1/2$) and $icnt$ is the iteration counter.
As the iteration progresses $r0$ becomes a very small negative quantity (approaching zero from below).
Weights and QoI are accumulated in the form $\id{wta}\gets\id{wta}+(r0\cdot\id{wta}+\id{wt})$.
Thereby the effective window length for accumulation grows roughly linearly in the iteration count.

\subsection*{Procedure $\id{Step}$}

Procedure $\id{Step}$ invokes the $\id{Step}$ procedure for $\id{WkHist\_Type}$ to advance the active walker one timestep or iteration and it maintains accumulated weights and QoI.
The QoI are obtained by a call to the $\id{Quan}$ procedure for $\id{WkHist\_Type}$, which in turn is just an interface to the $\id{Quan}$ procedure for $\id{Sys\_Config\_Type}$.
Here is the pseudo-code.

\begin{algorithmic}
\State $\id{wt0a}\gets\id{wt0a}+(r0\cdot\id{wt0a}+\id{wt})$ [Exponentially weighted accumulation of the active walker's weight $\id{wt}$ before the step.]
\State $\id{wk}\gets\id{wk}'$ [Invoking the $\id{WkHist\_Type}$ $\id{Step}$ procedure to advance the active walker.
This provides a new $(\id{cf},\id{wt},\id{wtb},\id{wti},\id{dti},\id{qvi})$.]
\State $\id{wta}\gets\id{wta}+(r0\cdot\id{wta}+\id{wt})$ [Exponentially weighted accumulation of the active walker's weight $\id{wt}$ after the step, but before rescaling.]
\State $qv\gets...$ [QoI from a call to the $\id{WkHist\_Type}$ $\id{Quan}$ procedure.]
\State $ic\gets...$ [Category index from a call to the $\id{WkHist\_Type}$ $\id{ICat}$ procedure.]
\State $\id{qa}(:,ic)\gets\id{qa}(:,ic)+(r0\cdot\id{qa}(:,ic)+\id{wt}\cdot qv)$ [Decaying accumulation of the active walker's QoI.]
\State $\id{wt}\gets\id{wt}\cdot\id{wt0a}/\id{wta}$ [Scale the active walker's weight to make the expected weight approximately constant.]
\end{algorithmic}

\subsection*{Procedure $\id{Strt}$}

Procedure $\id{Strt}$ maintains the pool of starters.
The motivation and an informal description was given earlier in the subsection on Starters and starter bias.
Here is the pseudo-code.

\begin{algorithmic}
\If{$\id{Random\_True}(\id{wt}/\id{wtb})$}
\State $l\gets...$ [Drawn from density $\rho(l)=2^{-l-1}$ for $0\leq l$, but mapped to $\id{nsql}-1$ if the draw exceeds that value.]
\State $\id{l0}\gets{...}$ [The least $l'$ for which $\id{isql}(l')=\id{isql}(l)$.]
\State $is\gets\id{isql}(\id{l0})-1$
\If{$\id{iwk0}+\id{nwk0}\leq is$}
\State [$is$ points to a location in the starter pool portion of the $\id{wks}$ array.]
\State $\id{wks}('is')\gets\id{wk}$ [The active walker is copied to $\id{wks}(:)$ at position $\hbox{modulo}(is,\id{nwks})$.]
\State $\id{wks}('is')\%\id{wt}\gets\max(\id{wt},\id{wtb})$ [The starter's weight is adjusted for surviving Russian Roulette.]
\State $\id{isql}(\id{l0})\gets is$ [Updating the $\id{isql}$ array.]
\Else
\State [The current active walker is skipped for the starter pool.]
\EndIf
\EndIf
\end{algorithmic}
As mentioned, $\id{Random\_True}(r)$ returns $\id{True}$ with probability $\id{median}(0,r,1)$.

The size $\id{nsql}$ of the $\id{isql}(0:\id{nsql}-1)$ array is chosen so that an index $\id{l0}$ near the top of $\id{isql}$ will appear seldom or never.

\subsection*{Procedure $\id{PBal}$}

Macroscopically population control or particle balance is maintained by the weight rescaling that is done in procedure $\id{Step}$ in order to maintain approximately constant average weight.
Microscopically there are the processes of splitting and Russian Roulette.
In the DMCD code the active walker is subject to these operations on every timestep or iteration.

A first version of the particle balance algorithm looks like this.

\begin{algorithmic}
\State[$\id{nwks}$ is the maximum size of the stack; $1\leq\id{nwks}$.]
\State[$\id{nwk0}$ is the actual size of the stack; $0\leq\id{nwk0}\leq\id{nwks}$.]
\State[$\id{wt}$ is the weight of the active walker.]
\State[$\id{wtb}$ is the base weight below which Russian Roulette may kill this walker.]
\State[$\id{rsp0}$ is the threshold for splitting, which is wanted if $\id{rsp0}\leq\id{wt}/\id{wtb}$.]
\State[$\id{rsp1}$ is a maintained upper bound on $\id{wt}/\id{wtb}$ that is applied if the stack is full.]
\State[We introduce some helper variables to clarify the logic.]
\State $b0\gets\id{Random\_True}(1-\id{wt}/\id{wtb})$ [The walker is terminated by Russian Roulette.]
\State $b1\gets\id{wt}/\id{wtb}<\id{rsp0}$ [The weight is below the splitting threshold.]
\State $b2\gets\id{nwk0}<\id{nwks}$ [The stack is not full.]
\If{b0}
\State $\id{Desc}$
[Accumulate descendant weighted QoI.]
\State $\id{Term}$
[Terminate the active walker and get a new one.]
\ElsIf{$(\lnot b0)\land(b1\lor\lnot b2)$}
\State [The active walker continues with possibly modified weight, compensating for the Russian Roulette if initially $\id{wt}<\id{wtb}$ and maintaining $\id{wt}/\id{wtb}\leq\id{rsp1}$ if the stack is full.]
\State $\id{wt}\gets\id{median}(\id{wtb},\id{wt},\id{rsp1}\cdot\id{wtb})$
\ElsIf {$(\lnot b0)\land b2\land\lnot b1$}
\State $\id{Desc}$
[Accumulate descendant weighted QoI.]
\State $\id{Spli}$
[Split the active walker in two. One continues and one goes onto the stack.]
\EndIf
\end{algorithmic}
In the presentation I provided complete guards, so that each guard expresses the condition under which the corresponding conditional branch is appropriate.

A complication arises in connection with descendant weighting due to the need to force splitting if some maximum number of steps is reached.
(Otherwise we might never obtain any history.)
The enhanced algorithm that is used in the code looks as follows.

\begin{algorithmic}
\State[$\id{nspf}$ is the maximum number of steps before a walker must be terminated or split.]
\State[$\id{ispf}$ is the current step counter, with $0\leq\id{ispf}\leq\id{nspf}$.]
\State[$b0$, $b1$ and $b2$ as before.]
\State $b3\gets\id{ispf}<\id{nspf}$ [The walker may continue without splitting.]
\If {$b0\lor\lnot(b2\lor b3)$}
\State $\id{Desc}$ [Accumulate descendant weighted QoI.]
\State $\id{Term}$ [Terminate the active walker and get a new one.]
\State $\id{ispf}\gets 0$
\ElsIf {$(\lnot b0)\land b3\land(b1\lor\lnot b2)$}
\State [The active walker continues with possibly modified weight.]
\State $\id{wt}\gets\id{median}(\id{wtb},\id{wt},\id{rsp1}\cdot\id{wtb})$
\State $\id{ispf}\gets\id{ispf}+1$
\ElsIf {$(\lnot b0)\land b2\land\lnot(b1\land b3)$}
\State $\id{Desc}$ [Accumulate descendant weighted QoI.]
\State $\id{Spli}$ [Split the active walker in two. One continues and one goes onto the stack.]
\State $\id{ispf}\gets 0$
\EndIf
\end{algorithmic}
Again I have included the complete guard for each branch as a statement of the condition under which the branch is appropriate.
It is worth verifying that the three guards are mutually exclusive and their union (disjunction) equals $\id{True}$.
This has already been seen for the case that $b3$ holds and one may complete the case analysis by looking next at the case $\lnot b3$.

\subsection*{Procedure $\id{Desc}$}

Descendant weighting involves accumulating quantities $\id{qvi}'\cdot(\id{wti}/\id{dti})$ in which $\id{qvi}'$ are accumulated QoI for a predecessor walker trajectory while $\id{wti}$ and $\id{dti}$ are accumulated quantities for the trajectory of the active walker (that is now being terminated or split).
It can only be done if the active walker has enough history on the stack.
Here is the pseudo-code.

\begin{algorithmic}
\State [Parameter $\id{nwkd}$ is the desired distance measured in splitting events.]
\If {$\id{nwkd}\leq\id{nwk0}$}
\State [The active walker has enough history on the stack.]
\State $is0\gets\id{iwk0}+\id{nwk0}-\id{nwkd}$ [Points to the relevant earlier stack location.]
\State $\id{qvi}'\gets...$ [Extracted from $\id{wks}('is0')$.]
\State $\id{qda}(:,:)\gets\id{qda}(:,:)+\id{qvi}'\cdot\id{wti}/\id{dti}$
\Else
\State [The active walker does not have enough history on the stack.]
\State [$\id{qda}(:,:)$ is unchanged.]
\EndIf
\end{algorithmic}

\subsection*{Procedure $\id{Term}$}

This procedure is invoked when the active walker is terminated (killed by Russian Roulette).
There are three cases.
If the stack is not empty then the new active walker is taken from the top of the stack; the position with index $is=\id{iwk0}+\id{nwk0}-1$.
If the stack is empty ($\id{nwk0}=0$) and a pool of starters is being maintained ($\id{isrc}=0$) then the new active walker is taken from the top of the pool; the position $is=\id{iwk0}+\id{nwks}-1$.
If the stack is empty and $\id{isrc}\not=0$ then the $\id{WkHist\_Type}$ $\id{Init}$ procedure is invoked to draw a walker from the defined (problem dependent) starter distribution.
(Recall that this is intended for development and testing purposes.)

Algorithmically it looks like this.
\begin{algorithmic}
\If {$\id{nwk0}=0\land\id{isrc}\not=0$}
\State $wk\gets...$ [The stack is empty and the new active walker comes from the $\id{Init}$ procedure.]
\Else
\State $wk\gets\id{wks}('\id{iwk0}+\id{nwk0}-1')$ [The new active walker comes from the $\id{wks}(:)$ array.]
\If {$\id{nwk0}=0$}
\State [The stack was empty, we obtained a starter.]
\State $\id{iwk0}\gets\id{iwk0}-1$
\Else
\State [We obtained a walker from the top of the stack.]
\State $\id{nwk0}\gets\id{nwk0}-1$
\EndIf
\EndIf
\end{algorithmic}

\subsection*{Procedure $\id{Spli}$}

The active walker is split into two, with one going onto the stack and the other continuing as the active walker.
A fraction $r0$ of the weight goes to the continuing active walker and a fraction $1-r0$ of the weight goes to the copy on the stack.
Here $r0=1/2$ for a regular split, but if the split was forced due to $\id{ispf}=\id{nspf}$ then $r0=1$.
(In this case the split is only for history purposes, not to diversify weights, hence all weight stays with the continuing walker.)
The history accumulators for the continuing active walker are reset to 0.

\section*{Experiments}

At first I was using the Schrödinger equation for a quadratic potential as a model system for experiments, but the inevitable timestep bias makes it difficult to interpret small effects.
For that reason the experiments for testing and development were finally all done using the Gaussian model implemented in the SysGaus module.

The default parameters in the code in the ancillary information are $nd=1$, $a0=1/2$, $a1=1/2$, $a2=1$, $a3=1$ and $a4=1e6$.
For these parameter values the exact stationary distribution has variance $0.5$ for regular weighting and $1.0$ for descendant weighting.

For a set of nine runs with $\id{nwks}=256$, $\id{rsp0}=13$, $\id{rsp1}=13$ (at the time these were automatically equal), $nwkd=8$ and using $5\times10^{10}$ iterations the results were
\begin{center}
\begin{tabular}{r l}
Means&[0.5000004 1.00006]\\
StdDevs&[4.3e-06 2.2e-04]
\end{tabular}
\end{center}
The measured deviations from the ideal averages [0.5 1.0] are well inside the standard deviations, so there is no measurable bias in these results.
At $\id{nwks}=64$, other parameters unchanged, there was still no measurable bias.

A slight bias might be visible at $\id{nwks}=16$, other parameters unchanged, but it is not obvious.
These are the results for a batch of 16 runs using again $5\times10^{10}$ iterations.
This was done using the exact code and input file that is in the ./anc/ directory.
\begin{center}
\begin{tabular}{r l}
Means&[0.4999932 1.00023]\\
StdDevs&[6.8e-06 2.0e-04]
\end{tabular}
\end{center}

Of course, tests on more interesting systems are required to affirm more strongly that the DMCD approach and the code are fundamentally sound.

\section*{Conclusions}

In this work it has been shown that DMC can be implemented using depth first traversal and a stack instead of a swarm.
This unifies the algorithmic treatment of particle transport Monte Carlo and DMC.

Certain algorithmic advantages are visible in the code.
The implementation of descendant weighting is very natural when using the history information that is in the stack.
The population control is simple enough to be done on every timestep.

The conjectured advantage of needing a much smaller stack than a typical swarm has been confirmed on a simple model system.

An issue that is not present in the breadth first, swarm based approach is the need to maintain a pool of starters.
A method for that was developed and implemented, and it too has held up in the tests for the model system.

Tests on more realistic quantum Monte Carlo applications will be required to validate the practical advantages of a smaller active memory footprint and of higher locality of computation.
In addition, future work should develop a better understanding of the finite stack-size bias and of the statistical quality of the starter pool.

\section*{Ancillary files (supplementary information)}

The ancillary information for this manuscript consists of one *.zip file that is stored in the {\tt ./anc/} directory associated with this Arxiv submission.
The zip archive contains the complete code that has been described here.

The code is set up to compile and run under the Fortran Package Manager (fpm).
Unpacking the zip archive produces a directory with the name {\tt DiffMC20260529-fpm}.
Within that directory there is a file {\tt fpm.toml} to control fpm, a file {\tt ArgsGaus} containing input data for a run, and subdirectories {\tt app} for the main program and {\tt src} for all the modules and submodules.

The code may be compiled and executed via the fpm command {\tt fpm run -- ArgsGaus}.
If fpm is not available then one can also concatenate the main program and all the modules and submodules, compile and link to obtain {\tt a.out}, and execute {\tt ./a.out ArgsGaus}.
I have used both Gnu gfortran and Intel ifx for this code.

Documentation within the code supplements the information in this manuscript.

\end{document}